\documentclass[
reprint,
superscriptaddress,
amsmath,amssymb,
aps,
floatfix
]{revtex4-2}
\usepackage[dvipdf]{graphicx}
\usepackage{bm}
\usepackage{txfonts}
\usepackage{braket}
\usepackage{color}      
\usepackage{ulem} 
\makeatletter
\newcommand{\figcaption}[1]{\def\@captype{figure}\caption{#1}}
\newcommand{\tblcaption}[1]{\def\@captype{table}\caption{#1}}
\makeatother

\begin{document}
\title{
Multipolar nematic state of nonmagnetic FeSe based on the DFT~+~$U$
}
\author{Takemi Yamada}\email{t-yamada@rs.tus.ac.jp}
\author{Takami Tohyama}
\affiliation{
Department of Applied Physics, Facility of Science, Tokyo University of Science, 6-3-1 Niijuku, Katsushika, Tokyo 125-8585, Japan
}
\date{\today}

\begin{abstract}
Clarifying the origin of nematic state in FeSe is one of urgent problems in the field of iron-based superconductivity. Motivated by the discovery of a nematic solution in the density-functional theory implemented by on-site Coulomb interaction (DFT~+~$U$) [npj Quantum Mater. {\bf 5}, 50 (2020)], we reexamine the $U$ dependence of electronic states in the nonmagnetic normal state of FeSe and perform full multipolar analyses for the nematic state. 
We find that with increasing $U$ the normal state experiences a topological change in the Fermi surfaces before the emergence of a nematic ground state. 
The resulting nematic ground state is a multipolar state having both antiferrohexadecapoles in the $E$ representation and ferromultipoles in the $B_2$ representation on each Fe site. 
Cooperative coupling between the $E$ and the $B_2$ multipoles in the local coordinate with the $D_{2d}$ point group will play an important role in the formation of the $d_{xz},~d_{yz}$ orbital-splitting nematic state not only in FeSe, but also in other iron pnictides.
\end{abstract}


\maketitle 
FeSe~\cite{Hsu2008,McQueen2009} is one of the most intensively studied iron-based superconductors~\cite{Mizuguchi-Takano2010,
Liu2015,
Bohmer-Kreisel2018,Shibauchi2020,
Kreisel2020} because of its complex and versatile ground state 
under pressure $P$~\cite{Sun2016,Terashima2015,Terashima2016a} and 
substitution of Se~\cite{Hosoi2016,Coldea2019}. 
Below the tetragonal-orthorhombic structural phase transition at $T_\mathrm{S}=90$~K, the electronic state of FeSe shows a behavior of the nematic state breaking the $C_{4}$ rotational symmetry whereas keeping translational symmetry without any magnetic ordering unlike other iron-based superconductors. 
Despite a tiny orthorhombicity~\cite{Bohmer2013}, the system exhibits a large band splitting associated with the orbital differentiation of $d_{xz},~d_{yz},~|E_{yz}-E_{xz}|=50$~meV~\cite{Shimojima2014}, which is too large to attribute to the lattice deformation. 
This strongly supports the electronic origin of the nematic state, being consistent with the enhancement of the nematic fluctuation when $T$ approaches $T_\mathrm{S}$ as observed in the nematic susceptibility~\cite{Bohmer2015,Hosoi2016} and the electronic Raman response~\cite{Massat2016,Zhang2020}. 
Therefore, the clarification of the nematic state is significant for 
the microscopic understanding of recent interesting phenomena, 
such as the orbital-selective correlation effect~\cite{Aichhorn2010,Liebsch-Ishida2010,Yin2011,Lanata2013} and the BCS-BEC crossover~\cite{Kasahara2014,Kasahara2016,Hanaguri2019,Kasahara2020}. 

The Fermi surface (FS) and low-energy band structures of FeSe have extensively been investigated~\cite{Terashima2014,Terashima2016b,
Nakayama2014,Maletz2014,Shimojima2014,Watson2015a,Suzuki2015,Zhang2015,Watson2016,Watson2017c,Pfau2019,Yi2019,
Subedi2008,Lohani2015} where the multiorbital compensated metal with Fe-$d$ orbitals is confirmed~\cite{Shibauchi2020,Kreisel2020}.
In the normal (nonnematic) state, two hole FSs (2h-FSs) around $\Gamma$ and two electron FSs (2e-FSs) around M have commonly been reported, but their size is extremely small only occupied 2-3 \% in the Brillouin zone (BZ). 
These small FSs and the low-energy band structure have not still been reproduced by 
the density-functional theory (DFT)~\cite{Subedi2008}, 
DFT~+~$U$~\cite{Lohani2015}, the dynamical mean field theory (DMFT)~\cite{Aichhorn2010,Liebsch-Ishida2010,Yin2011,Mandal2014}, and the quasiparticle self-consistent $GW$~\cite{Tomczak2012}. 
Several studies based on adjusted models to reproduce the low-energy bands of angle-resolved photoemission spectroscopy (ARPES)~\cite{Suzuki2015} can explain the enhancement of orbital and magnetic fluctuations in the $T$-$P$ phase
~\cite{Yamakawa2016,Yamakawa-Kontani2017,Ishizuka2018}.

As for the nematic state, several FSs have been reported by 
the Shubnikov-de Haas~\cite{Terashima2014,Terashima2016b} and ARPES
experiments~\cite{Nakayama2014,Maletz2014,Shimojima2014,Watson2015a,Suzuki2015,Zhang2015,Watson2016,Watson2017c,Pfau2019,Yi2019} 
where a single hole FS (1h-FS) near $\Gamma$ is common whereas it is still unsettled whether the electron FS near M is a single (1e-FS) or two. 
The sign change in the orbital splitting at $\Gamma$ and M points in the BZ has been observed~\cite{Watson2015a,Suzuki2015}, whose origin and mechanism have been discussed~\cite{Su2015,Onari2016,Xing2018,Kang2018}.
The recent DFT study~\cite{Long2020} has provided a new nematic ground state with the $E_u$ irreducible representation of the $D_{4h}$ symmetry, which contains 1e-FS and additional hybridization between $d_{xy}$ and $d_{xz},~d_{yz}$ orbitals~\cite{Steffensen2021,Rhodes2021}. 
Although this nematic state seems to explain the recent experiment~\cite{Yi2019}, 
it is unclear how the nematic state is reached from the well-known three hole FSs (3h-FSs) of the DFT normal state~\cite{Lohani2015,Long2020}. 
Therefore, a systematic investigation of the normal state 
on the verge of nematic ordering and a detailed multipolar analysis in the nematic state are highly desirable.

In this Letter, we examine the $U$ dependence of the electronic states of FeSe by the DFT~+~$U$ method and find a topological change in FSs before 
a nematic order occurs. 
The resulting nematic ground state is found to be a multipolar state having both antiferrohexadecapoles in the $E$ representation and ferromultipoles in the $B_2$ representation on each Fe site with the locally $D_{2d}$ point group. This coexistence indicates that cooperative coupling between the $E$ and the $B_2$ multipoles can be a source of the formation of the $d_{xz},~d_{yz}$ orbital-splitting nematic state in FeSe and related materials. 

We have performed the DFT~+~$U$ calculation~\cite{SM} in the first-principles code WIEN2k~\cite{Blaha2020} 
where the Coulomb interaction $U$ for $d$ electrons in the muffin-tin~(MT) radius $R_{\rm MT}^{\nu}$ with atomic sites $\nu=$Fe1,~Fe2 in the unit cell is introduced. 
The DFT~+~$U$ correction energy consists of total occupation number of $d$ electrons within $R_{\rm MT}^{\nu},~n^{d,\nu}=\sum_{m\sigma}n_{mm}^{\sigma,\nu}$, and the density-matrix $n_{mm'}^{\sigma,\nu}$ as explicitly shown in Ref.~\cite{SM}. 
Hereafter we drop the spin index $\sigma$ and use $n_{mm'}^{\uparrow,\nu}=n_{mm'}^{\downarrow,\nu}=n_{mm'}^{\nu}$ and $n_{mm}^{\nu}=n_{m}^{\nu}/2$ due to the nonmagnetic situation throughout the paper~\cite{SM}. 
By solving the Kohn-Sham equation self-consistently, the band energy $\varepsilon_{\bm{k}n}$ with wave-vector $\bm{k}$ and band-index $n$ is obtained for any given $U$ where the effective $+U$ potential acting on the atomic basis 
$v_{mm'}^{\nu}$~\cite{Shick1999,Tran2008} is given by 
\begin{align}
&v_{mm'}^{\nu}=\delta_{mm'}\frac{U}{2}(1-n_{m}^{\nu})+(1-\delta_{mm'})(-Un_{mm'}^{\nu}),
\label{eq:vmm}
\end{align}
where the first (second) term in Eq.~(\ref{eq:vmm}) is proportional to the diagonal (off-diagonal) density matrix $n_{mm}^{\nu}$~($n_{mm'}^{\nu}$). 
All the technical details are presented in Ref.~\cite{SM}.

\begin{figure}[t]
\centering
\includegraphics[width=7.5cm]{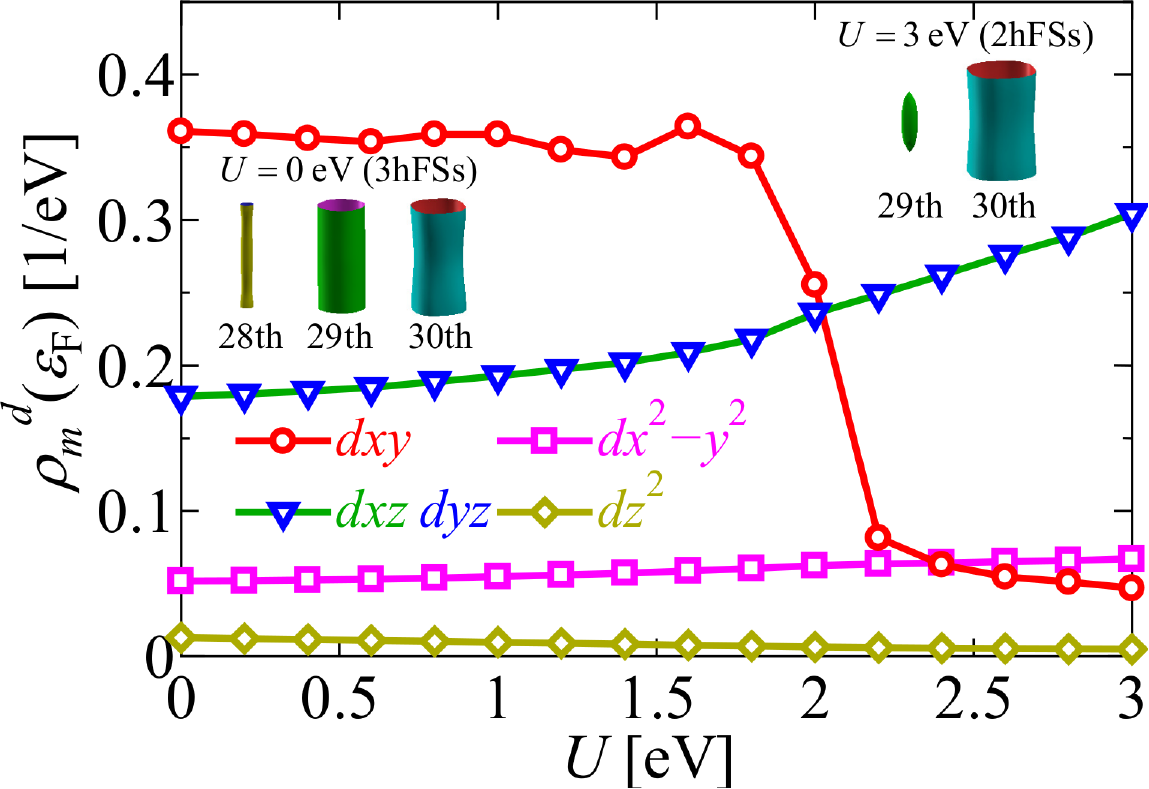}
\vspace{-0.2cm}
\caption{(Color online) 
$U$ dependence of pDOS $\rho_{m}^{d}(\varepsilon_{\rm F})$. 
The inset shows the 3h-FSs for $U=0$~eV (left center) and the 2h-FSs for $U=3$~eV (right top). 
}
\label{Fig01}
\end{figure}

First, we investigate the $U$ dependence of the normal electronic state in FeSe. 
Figure~\ref{Fig01} shows the partial density-of-states (pDOS) $\rho_{m}^{d}(\varepsilon_{\rm F})$ at the Fermi energy $\varepsilon_{\rm F}$. 
When $U=0$~eV, 3h-FSs consisting of the 28th, 29th, and 30th bands are obtained as shown in the inset of Fig. \ref{Fig01}, being similar to the previous DFT results~\cite{Subedi2008,Lohani2015}. 
With increasing $U$, $\rho_{m}^{d}(\varepsilon_{\rm F})$ of $d_{xy}$ orbital drops at $U\simeq 2$~eV (=$U_{\rm LT}$), 
whereas, in turn, that of $d_{xz},~d_{yz}$ increases gradually. 
Since the 28th band constructing the most inner hole FS originates from the $d_{xy}$ orbital, it falls below $\varepsilon_{\rm F}$ and the FS vanishes at a Lifshitz transition point $U_{\rm LT}$ where the 3h-FSs change to the 2h-FSs from the 29th and 30th bands as shown in the inset of Fig.~\ref{Fig01}~\cite{memo-2h-FSs}. 
This change is caused by $v_{U}(\bm{r})$ that induces orbital-dependent energy shifts, being proportional to $1-n_{m}^{\nu}$ as shown in the first term of Eq. (\ref{eq:vmm}). We can find that the occupied number $n_{m}^{d}$ in $d_{xy}$ and $d_{z^2}$ at $U=0$~eV, which is larger than in $d_{xz},~d_{yz}$ and $d_{x^2-y^2}$, increases further with increasing $U$, resulting in the negatively large value of $1-n_{m}^{\nu}$ in $d_{xy}$~\cite{DFT-particle-number}.

In contrast to the h-FSs, the e-FSs from the 31st and 32nd bands with $d_{xz},~d_{yz}$ orbitals at M and A points
do not undergo any topological change but, instead, 
$\varepsilon_{\bm{k}n}$ of these bands approaches $\varepsilon_{\rm F}$ monotonically 
with increasing $U$. This manifestation of the $d_{xz},~d_{yz}$ orbitals near $\varepsilon_{\rm F}$ as evidenced by the 2h-FSs and e-FSs will triggers the formation of a nematic state mentioned below.

\begin{figure}[t]
\centering
\includegraphics[width=8.5cm]{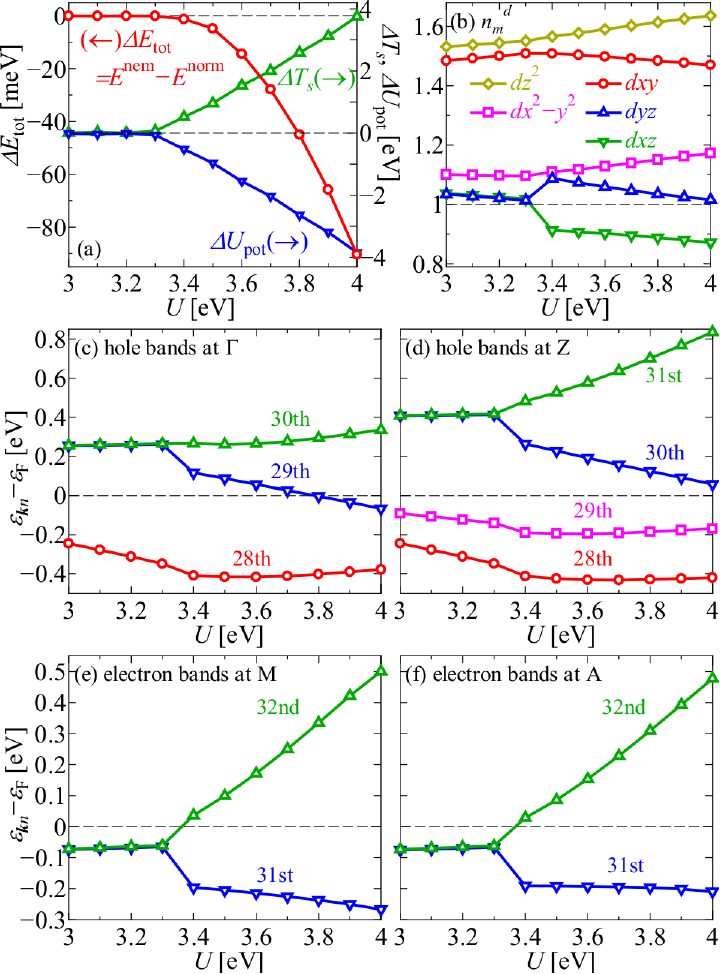}
\vspace{-0.25cm}
\caption{(Color online) 
(a) Energy differences of the total energy $\Delta E_{\rm tot}$ (left axis), and kinetic and potential energies $\Delta T_{s},~\Delta U_{\rm pot}$ (right axis) and (b) $n_{m}^{d}$.
(c)$-$(f) The obtained band energy $\varepsilon_{\bm{k}n}-\varepsilon_{\rm F}$ around $\varepsilon_{\rm F}$ for $U=3$ to 4~eV.
}
\label{Fig02}
\end{figure}

Next we calculate a nematic solution by preconditioning the initial charge density~\cite{SM} 
as was performed in the previous pseudopotential calculation~\cite{Long2020}. 
Figure~\ref{Fig02} (a) shows the total energy difference between the normal and the nematic states $\Delta E_{\rm tot}=E_{\rm tot}^{\rm nem}-E_{\rm tot}^{\rm normal}$ as a function of $U$ together with the kinetic-energy and potential-energy differences $\Delta T_{s}$ and $\Delta U_{\rm pot}$, where $E_{\rm tot}=T_{s}+U_{\rm pot}$ is calculated by the total energy formula of the all-electron method~\cite{Weinert1982} under the Virial theorem $2T_{s}+U_{\rm pot}=0$. 
With increasing $U$ more than $U=3.4$~eV, $\Delta E_{\rm tot}<0$ is realized with $\Delta T_{s}>0$ and $\Delta U_{\rm pot}<0$ together with the occupied number splitting between $d_{xz}$ and $d_{yz}$ orbitals as shown in Fig.\ref{Fig02}(b). 
The energy gain is on the order of $O(10^{1-2}{\rm meV})$, which basically agrees with the previous pseudopotential DFT result~\cite{Long2020}. 

The $U$ dependence of $\varepsilon_{\bm{k}n}-\varepsilon_{\rm F}$ near $\varepsilon_{\rm F}$ 
at $\Gamma$ and Z [M and A] points is shown in Figs.~\ref{Fig02}(c) and \ref{Fig02}(d) [Figs.~\ref{Fig02}(e) and \ref{Fig02}(f)], respectively. 
The $d_{xz}$- and $d_{yz}$-orbital bands, whose number is denoted in the figures, split due to the nematic state transition.
Since the 32nd band in M and A points rises above $\varepsilon_{\rm F}$, 2e-FSs in the nonnematic state change to 1e-FS in the nematic state. 
Combining with the change from 2h-FSs to 1h-FS at $\Gamma$ point above $U$=3.7~eV, we find the number of FSs consistent with the experiment~\cite{Yi2019} and the previous DFT result~\cite{Long2020}.

\begin{table}[t]
\caption{
The definition and notations of multipole $X_{\alpha}$ denoted by quadrupole $O_{\Gamma_{\gamma}}$ and hexadecapole $Q_{4_{\Gamma_{\gamma}}}$ together with the irreducible representations (IRRs) in $D_{2d}$ symmetry~\cite{Kusunose2008,Hayami2018} at each Fe site.  $T_{kq}^{(c,s)}~(T_{kq})$ is the tesseral (spherical) tensor operator with a relation $T_{kq}^{(c,s)}=\frac{(-1)^q}{\sqrt{2}}(T_{kq}^{\dagger}\pm T_{kq})$.
}\label{table01}
\centering
\begin{tabular}{cccc}
\hline \rule{0pt}{4mm}
IRR & $X_{\alpha}$ & Tesseral tensor representation & $(x,y,z)$ representation \\[2pt] \hline\hline \rule{0pt}{4mm}
$A_{1}$ & $O_{u}$   & $T_{20}$    & $\frac{1}{2}(3z^2-r^2)$ \\[2pt] \rule{0pt}{4mm}
          & $Q_{4}$   & $\sqrt{\frac{5}{12}}T_{44}^{(c)}+\sqrt{\frac{7}{12}}T_{40}$ & $\frac{5\sqrt{21}(x^4+y^4+z^4-\frac{3}{5}r^4)}{12}$ \\[2pt] \rule{0pt}{4mm}
          & $Q_{4u}$ & $-\sqrt{\frac{7}{12}}T_{44}^{(c)}+\sqrt{\frac{5}{12}}T_{40}$ & $\frac{7\sqrt{15}[2z^4-x^4-y^4-\frac{6}{7}r^2(3z^2-r^2)]}{12}$ \\[2pt] \rule{0pt}{4mm}
$A_{2}$ & $Q_{4\alpha,z}$ & $-T_{44}^{(s)}$ & $\frac{\sqrt{35}}{2}xy(x^2-y^2)$ \\[2pt] \rule{0pt}{4mm}
$B_{1}$ & $O_{v}$   & $T_{22}^{(c)}$ & $\frac{\sqrt{3}}{2}(x^2-y^2)$ \\[2pt] \rule{0pt}{4mm}
          & $Q_{4v}$ & $-T_{42}^{(c)}$ & $\frac{7\sqrt{5}[x^4-y^4-\frac{6}{7}r^2(x^2-y^2)]}{4}$ \\[2pt] \rule{0pt}{4mm}
$B_{2}$ & $O_{xy}$ & $T_{22}^{(s)}$ & $\sqrt{3}xy$ \\[2pt] \rule{0pt}{4mm}
          & $Q_{4\beta,z}$ & $T_{42}^{(s)}$ & $\frac{\sqrt{5}}{2}xy(7z^2-r^2)$ \\[2pt] \rule{0pt}{4mm}
$E$ & $O_{zx},O_{yz}$ & $T_{21}^{(c)},T_{21}^{(s)}$ & $\sqrt{3}zx,\sqrt{3}yz$ \\[2pt] \rule{0pt}{4mm}
      & $Q_{4\alpha,x}$ & $-\sqrt{\frac{1}{8}}T_{43}^{(s)}-\sqrt{\frac{7}{8}}T_{41}^{(s)}$ & $\frac{\sqrt{35}}{2}yz(y^2-z^2)$ \\[2pt] \rule{0pt}{4mm}
      & $Q_{4\alpha,y}$ & $-\sqrt{\frac{1}{8}}T_{43}^{(c)}+\sqrt{\frac{7}{8}}T_{41}^{(c)}$ & $\frac{\sqrt{35}}{2}zx(z^2-x^2)$ \\[2pt] \rule{0pt}{4mm}
      & $Q_{4\beta,x}$  & $  \sqrt{\frac{7}{8}}T_{43}^{(s)}-\sqrt{\frac{1}{8}}T_{41}^{(s)}$ & $\frac{\sqrt{5}}{2}yz(7x^2-r^2)$ \\[2pt] \rule{0pt}{4mm}
      & $Q_{4\beta,y}$  & $-\sqrt{\frac{7}{8}}T_{43}^{(c)}-\sqrt{\frac{1}{8}}T_{41}^{(c)}$ & $\frac{\sqrt{5}}{2}zx(7y^2-r^2)$ \\[2pt] 
\hline \rule{0pt}{4mm}
\end{tabular}
\end{table} 

\begin{figure}[t]
\centering
\includegraphics[width=8.5cm]{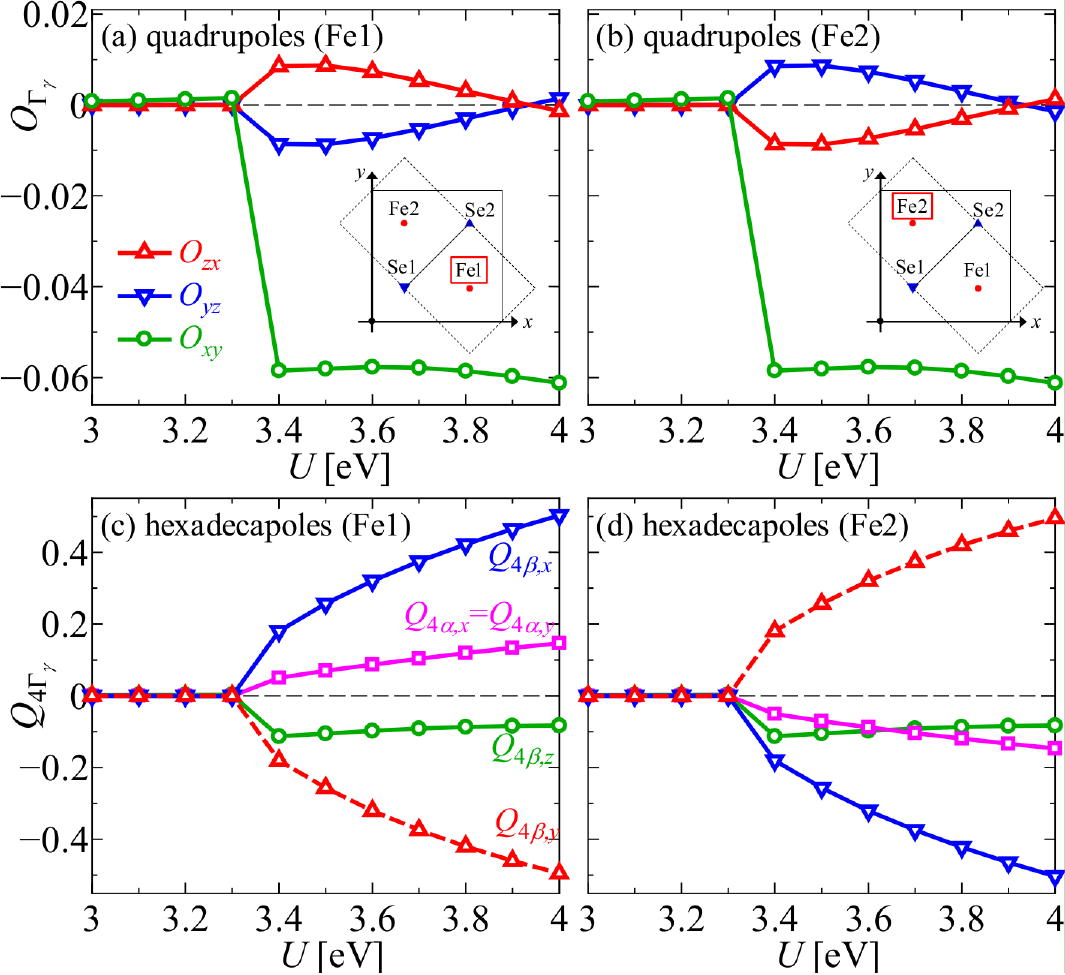}
\vspace{-0.25cm}
\caption{(Color online) 
The quadrupole and hexadecapole moments (a) and (b) $O_{\Gamma_\gamma}$ and (c) and (d) $Q_{4\Gamma_\gamma}$ 
on Fe1 [(a) and (c)] and Fe2 [(b) and (d)] sites as a function of $U$. In the insets of (a) and (b), atom sites in the unit cell are depicted with $x,y$-axes.
}
\label{Fig03}
\end{figure}

The order parameters of the nematic state obtained by DFT~+~$U$ have been discussed in the previous study~\cite{Long2020}, where only finite off-diagonal density matrix elements are taken into account. 
Here, we derive all the active multipole moments in the present system more generally~\cite{Kusunose2008,Hayami2018}. 
Without the spin-orbit interaction, the multipole operator can be regarded as a power series expansion of the rank $k$ of orbital angular momentum operator $\bm{\ell}=(\ell_{x},\ell_{y},\ell_{z})$. 
Only the even rank multipoles, i.e., quadrupoles ($k=2$) and hexadecapoles ($k=4$), become finite for the $d$ electron basis in each Fe site due to the time reversal symmetry. 
These multipoles are classified by the irreducible representations \{$A_1$, $A_2$, $B_1$, $B_2$, $E$\} at the Fe site with $D_{2d}$ symmetry, which are summarized in Table \ref{table01}.

We calculate all the multipole moments listed in Table \ref{table01} from the density matrix $n_{mm'}^{\nu}$, where all the multipole operators are normalized as ${\rm Tr}[X_{\alpha}X_{\beta}]=\delta_{\alpha\beta}$. 
We note that a similar approach has been performed on the magnetic multipole order in the actinide dioxides~\cite{Suzuki2013,Suzuki2018}.
We find that the multipoles directly related to the nematic order are those in the $B_2$ and $E$ representations. 
Figures~\ref{Fig03}(a) and \ref{Fig03}(c) [\ref{Fig03}(b) and \ref{Fig03}(d)] show the $U$ dependence of the quadrupoles $O_{xz},~O_{yz},~O_{xy}$ and the hexadecapoles $Q_{4\beta, z},~Q_{4\alpha, x},~Q_{4\alpha, y},~Q_{4\beta, x},~Q_{4\beta, y}$ at the Fe1 [Fe2] site, respectively. 
Above $U=3.4$~eV, the $B_{2}$ quadrupole moment for $O_{xy}$ at the two Fe sites becomes negative as seen in Figs.~\ref{Fig03}(a) and \ref{Fig03}(b) and the similar behavior is obtained for the $B_{2}$ hexadecapole $Q_{4\beta, z}$ as seen in Figs.~\ref{Fig03}(c) and \ref{Fig03}(d). 
This behavior corresponds to the emergence of a ferro nematic order associated with the orbital differentiation between $d_{xz}$ and $d_{yz}$. 
On the other hand, the $E$ hexadecapole moments for $Q_{4\alpha, x},~Q_{4\alpha, y},~Q_{4\beta, x}$, and $Q_{4\beta, y}$ are more than ten times larger than the $E$ quadrupole moments for $O_{xz},O_{yz}$. It is also interesting to note that $Q_{4\beta, x}>0,~ Q_{4\beta, y}<0$, and $Q_{4\alpha, x(y)}>0$ 
at Fe1 but opposite signs at Fe2 as shown in Figs~.\ref{Fig03}(c) and \ref{Fig03}(d). 
Namely, the $E$-type order parameter at Fe1 (Fe2) is written as $\pm(aQ_{4\alpha}^{E}+bQ_{4\beta}^{E}+cO^{E})$, 
where 
$Q_{4\alpha(\beta)}^{E}=\frac{1}{\sqrt{2}}(Q_{4\alpha(\beta),x}\pm Q_{4\alpha(\beta),y})$ and $O^{E}$=$\frac{1}{\sqrt{2}}(O_{zx}-O_{yz})$ 
with $a^2+b^2+c^2=1$. 
This result indicates that the antiferro ordering of the $E$ multipoles with opposite values at two Fe sites 
coexists with the ferro ordering of the $B_{2}$ multipoles with the same values at two sites.

\begin{figure*}[t]
\centering
\includegraphics[width=17.5cm]{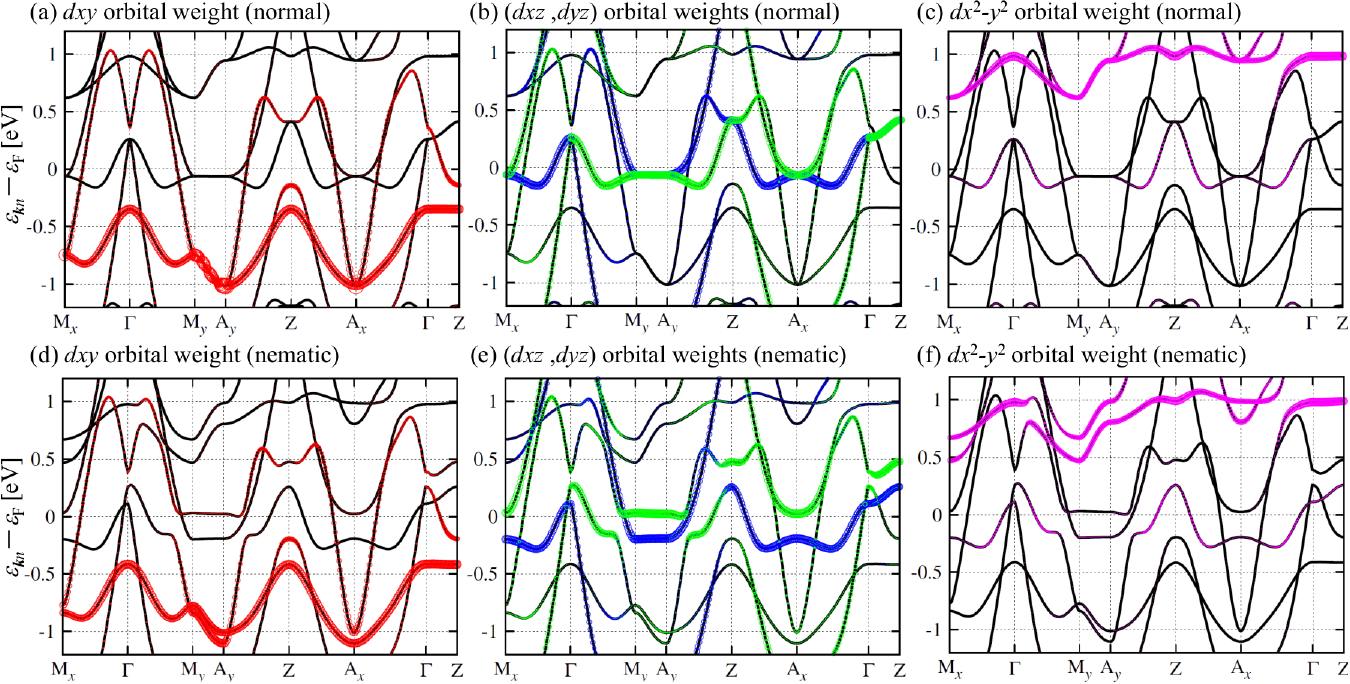}
\vspace{-0.3cm}
\caption{(Color online) 
Band structures for (a)$-$(c) normal state ($U=3.3$~eV) and (d)$-$(f) nematic state ($U=3.4$~eV) 
with the orbital weight of 
(a) and (d) $d_{xy}$, 
(b) and (e) $d_{xz}$(green), $d_{yz}$(blue), and 
(c) and (f) $d_{x^2-y^2}$ 
along the high-symmetry line in the BZ, where 
${\rm M}_{x}[{\rm M}_{y}]=(\frac{\pi}{a},\frac{\pi}{a},0)[(\frac{\pi}{a},-\frac{\pi}{a},0)]$
and 
${\rm A}_{x}[{\rm A}_{y}]=(\frac{\pi}{a},\frac{\pi}{a},\frac{\pi}{c})[(\frac{\pi}{a},-\frac{\pi}{a},\frac{\pi}{c})]$ and $a,c$ are the lattice constants.
}
\label{Fig04}
\end{figure*}

Such the coexistence of $E$- and $B_2$-type multipoles can be understood from the phenomenological intermultipole coupling theory~\cite{Kusunose2008,Kusunose-Kuramoto2001} at finite temperature $T$, 
where the Ginzburg-Landau free energy with the mean-field (MF) approximation can be expanded by the multipole moment $X_{\alpha}$ around the nematic transition as given by $F_{\rm MF}=F_{\rm MF}^{\rm (2nd)}+F_{\rm MF}^{\rm (3rd)}+\cdots$. 
Here, $F_{\rm MF}^{\rm (3rd)}=-\frac{T}{3!}\sum_{\alpha\beta\gamma}g_{\alpha\beta\gamma}X_{\alpha}X_{\beta}X_{\gamma}$, 
where $X_{\alpha}$=$O_{\Gamma_{\gamma}}$ or $Q_{4\Gamma_{\gamma}}$ and 
$g_{\alpha\beta\gamma}$ is the symmetric constant defined as $g_{\alpha\beta\gamma}=\frac{1}{2d}{\rm Tr}\left[(X_{\alpha}X_{\beta}+X_{\beta}X_{\alpha})X_{\gamma}\right]$~\cite{MP-coupling} with matrix dimension $d=5$. 
The coupling terms among $O_{xy},~Q_{4\beta,z}$ and $E$ multipoles are explicitly given by 
$F_{\rm MF}^{\rm (3rd)}=
-Tc_{1}O_{xy}\left(-\frac{4}{11}O_{xz}O_{yz}-\frac{7}{11}Q_{4\alpha,x}Q_{4\alpha,y}+Q_{4\beta,x}Q_{4\beta,y}\right)
-Tc_{2}Q_{4\beta,z}\left(O_{xz}O_{yz}-\frac{7}{8}Q_{4\alpha,x}Q_{4\alpha,y}+Q_{4\beta,x}Q_{4\beta,y}\right)$, 
where $c_1(c_2)=\tfrac{11}{84}\sqrt{\tfrac{15}{14}}(\frac{1}{21}\sqrt{\frac{10}{7}})$ 
and the second and third terms in both parentheses correspond to the coupling between the $E$ multipoles and are negative 
whereas the coefficients of both parentheses including the $B_{2}$ multipoles are positive as shown in Fig.~\ref{Fig03}. 
Therefore, $F_{\rm MF}^{\rm (3rd)}$ becomes negative as a whole stabilizing the coexistence state.

Finally, we discuss the band structure and orbital components of the normal ($U=3.3$~eV) and nematic ($U=3.4$~eV) states as shown in Figs.~\ref{Fig04}(a)$-$\ref{Fig04}(c) and \ref{Fig04}(d)$-$\ref{Fig04}(f), respectively. 
It is clearly observed that the degenerated $d_{xz}$ and $d_{yz}$ bands near 0.25~eV (0.5~eV) at the $\Gamma$~(Z) point split in the nematic state, corresponding to the ordering of $O_{xy}$ and $Q_{4\beta, z}$ as seen in Fig.~\ref{Fig03}. 
Band splitting due to the same mechanism is also realized in the 31st and 32nd electron bands near 0~eV along the M-A direction, 
which consist of linear combinations of $d_{xz}$ and $d_{yz}$ orbitals without $k_z$ dependence. 
This leads to a splitting of the peak structure in pDOS $\rho_{m}^{d}(\varepsilon_{\rm F})$ of the $d_{xz}$ and $d_{yz}$ orbitals~\cite{SM}. 
In the nematic state as shown in Fig.~\ref{Fig03}, the $E$ hexadecapoles become finite, 
which gives rise to the significant mixings among the $|\ell_z|=2$ orbitals ($d_{xy}$ and $d_{x^2-y^2}$) and $|\ell_z|=1$ orbitals ($d_{xz}$ and $d_{yz}$)\cite{E-quadrupole}. 
The mixing with $d_{x^2-y^2}(d_{xy})$ having a large weight above (below) $\varepsilon_{\rm F}$ 
pushes down (up) the energy level of the mixed partner $d_{yz}(d_{xz})$. 
Such a mixing effect is remarkable in the M$_y$-$\Gamma$ and A$_{y}$-Z directions as seen in Figs.~\ref{Fig04}(a) and \ref{Fig04}(d): A mixing gap is formed near $-0.1$~eV in their directions. 
On the other hand, the band for the $d_{xy}$ orbital near $-0.2$~eV in the M$_x$-$\Gamma$ and A$_x$-Z directions remains almost unchanged. 
This one-side gap opening along the M$_x$-$\Gamma$-M$_y$ direction, 
which is consistent with the experiment~\cite{Yi2019}, 
is critically important for the origin of the nematic state because it inherently requires the presence of the $E$ multipoles as pointed out in this Letter~\cite{sign-reversing}.

The orbital dependent correlation effect due to Hund's coupling $J$, ``Hund's metal'' behavior, has been discussed in FeSe, where the pressure- and correlation-driven Lifshitz transitions~\cite{Skornyakov2018,SL2019}, and the enhancement of compressibility with a charge instability~\cite{Arribi-Medici2018,Chatzieleftheriou2020} have been obtained for the similar $U$ values of the present nematic transition. 
Therefore, it will be important to extend the present method to a strongly correlated theory incorporating the properties of Hund's metal and to clarify the relation between the nematic states obtained here and the Hund's metal phenomena~\cite{Skornyakov2018,SL2019,Arribi-Medici2018,Chatzieleftheriou2020}, which is, however, beyond the scope of the present Letter.

To summarize, we have studied the nonmagnetic normal and nematic states of the iron-based superconductor FeSe by using the DFT~+~$U$ method with the multipole analyses. 
The effect of $U$ on the normal state generates a topological change in FSs from 3h-FSs to 2h-FSs, 
leading to a change in the dominant orbital near $\varepsilon_{\rm F}$ from $d_{xy}$ to $d_{xz},~d_{yz}$. 
As a result, the multipolar nematic state with the $E$ antiferrohexadecapoles accompanying the $B_2$ ferromultipoles 
has been obtained without any assumption of the order parameters, 
giving rise to both of the $d_{xz}$-$d_{yz}$ orbital splitting at $\Gamma$ and the $d_{xy}$-$(d_{xz},~d_{yz})$ orbital mixing around M and A points. 
From phenomenological analysis, we have found that the intermultipole coupling of $B_2$ and $E$ multipoles on each Fe site can explain the energy gain larger for the coexisting order than for the quadupole $O_{xy}$ order alone. 
This multipolar mechanism for the formation of nematic state will be applicable 
not only to FeSe but also to other iron pnictides where the degenerated $d_{xz}$,$d_{yz}$ orbitals play a crucial role.

\begin{acknowledgments}
This work was supported by the ``Quantum Liquid Crystals'' No. JP19H05825 KAKENHI on Innovative Areas from JSPS of Japan.
\end{acknowledgments}

\bibliography{FeSe_nematic-DFT}

\begin{thebibliography}{72}
\bibitem [{mem()}]{memo-2h-FSs}%
  \BibitemOpen
  \href@noop {} {}\bibinfo {note} {Here we also check
  $\varepsilon_{\bm{k}n}-\varepsilon_{\rm F}$ around $\varepsilon_{\rm F}$ for
  the hole (electron) bands at $\Gamma$ and Z (M and A) points as a function
  $U$, and we obtain a continuous tracking of the connection from the 3h-FSs
  and 2h-FSs together with a critical interaction of the Lifshitz transition
  $U_{\rm LT}$, which is one of the new results for the present Letter in
  contrast to the previous DFT+$U$ study~\cite{Lohani2015} where the 2h-FSs
  were obtained only for several large values of $U$.}\BibitemShut {Stop}%
\bibitem [{DFT()}]{DFT-particle-number}%
  \BibitemOpen
  \href@noop {} {}\bibinfo {note} {Throughout the calculation, total $d$
  occupation number $n^{d,\nu}$ is almost unchanged from $n^{d,\nu}$=6.17-6.19
  and the total valence electron number $n_{\rm v}$ is strictly fixed $n_{\rm
  v}=60$~per two (FeSe) in the unit cell where 28 electrons are attributed to
  Fe-$3p,3d,4s$ and 32 electrons to Se-$3d,4s,4p$.}\BibitemShut {Stop}%
\bibitem [{MP-()}]{MP-coupling}%
  \BibitemOpen
  \href@noop {} {}\bibinfo {note} {Here we normalize the multipoles as
  $\frac{1}{d}{\rm Tr}[X_{\alpha}X_{\beta}]=\delta_{\alpha\beta}$ with $d=5$
  unlike the calculation of the density matrix $n_{mm'}^{\nu}$.}\BibitemShut
  {Stop}%
\bibitem [{E-q()}]{E-quadrupole}%
  \BibitemOpen
  \href@noop {} {}\bibinfo {note} {Here we note that the $E$ quadrupoles
  $O_{zx},O_{yz}$ also induce the similar mixings but their hybridizations are
  partial and incomplete, since in $|m\rangle$ orbital basis with
  $m=-2,\cdots,+2$,~$O_{\Gamma_{\gamma}}$ with $\Gamma_{\gamma}=zx,~yz$ has
  only finite elements as $\langle m|O_{\Gamma_{\gamma}}|m\pm 1\rangle$, while
  the hexadecapoles $Q_{4\Gamma_{\gamma}}$ with $\Gamma_{\gamma}=\alpha,x(y)$
  and $\beta,x(y)$ have finite elements as $\langle m|Q_{4\Gamma_{\gamma}}|m\pm
  3\rangle$ in addition to $\langle m|Q_{4\Gamma_{\gamma}}|m\pm
  1\rangle$.}\BibitemShut {Stop}%
\end{thebibliography}

\clearpage
\newpage
\clearpage 
\setcounter{equation}{0}%
\setcounter{figure}{0}%
\setcounter{table}{0}%
\renewcommand{\thetable}{S\arabic{table}}
\renewcommand{\theequation}{S\arabic{equation}}
\renewcommand{\thefigure}{S\arabic{figure}}
\renewcommand{\thesection}{S\arabic{section}}
\newtheorem{proposition}{Proposition}
\onecolumngrid

\begin{center}
{\large{\bf Supplemental Material for\\ ``Multipolar nematic state of nonmagnetic FeSe based on the DFT~+~$U$''}}\\
\vspace{5mm}
T. Yamada and T. Tohyama
\vspace{5mm}
\end{center}

\twocolumngrid

\section{Computatinal details}
Here we provide the details of the DFT calculation based on WIEN2k~\cite{Blaha2020}, which is the all-electron first-principles code, 
where the basis functions are expanded by the relativistic full-potential augmented plane wave (FLAPW) and/or APW + local orbitals depending on the core and valence states of each atom in the unit-cell. 
We employ the generalized gradient approximation for the exchange-correlation potential of the PBE-GGA potential~\cite{PBE-GGA1996} 
and the DFT~+~$U$ method within the self-interaction correction~\cite{Anisimov1993,Shick1999}. 
The spin-orbit interaction (SOI) is neglected in this study to simplify the multipole analysis, 
since its effect on the nematics state is weak as shown in Fig.~7 of Ref.~\cite{Long2020} and 
does not give any serious problem on the main results in the present study. 

The Coulomb interaction $U$ for $d$ electrons in the muffin-tin radius $R_{\rm MT}^{\nu}$ with atomic sites $\nu$=Fe1,Fe2 in the unit cell is introduced, where the DFT~+~$U$ correction energy is given by, 
\begin{align}
E_{U}=\frac{U}{2}\sum_{\nu}\left(n^{d,\nu}-\sum_{mm'}\sum_{\sigma}n_{mm'}^{\sigma,\nu}n_{m'm}^{\sigma,\nu}\right),
\end{align}
where $n^{d,\nu}$ is total occupation number of $d$ electrons within $R_{\rm MT}^{\nu}$ as $n^{d,\nu}=\sum_{m\sigma}n_{mm}^{\sigma,\nu}$ and the density matrix $n_{mm'}^{\sigma,\nu}$. Hereafter, we drop the spin index $\sigma$ and use $n_{mm'}^{\uparrow,\nu}$=$n_{mm'}^{\downarrow,\nu}$=$n_{mm'}^{\nu}$ and $n_{mm}^{\nu}$=$n_{m}^{\nu}/2$, since only the nonmagnetic state is studied. 

In WIEN2k, the Kohn-Sham (KS) equation for single electron, depending on the core and valence states of each atom in the unit-cell, is explicitly written by 
\begin{align}
\left(-\frac{1}{2}\bm{\nabla}^2+v^{\rm KS}_{{\rm eff}}(\bm{r})+v_{U}(\bm{r})\right)\psi_{\bm{k}n}(\bm{r})=\varepsilon_{\bm{k}n}\psi_{\bm{k}n}(\bm{r}),\label{eq:KS}
\end{align}
where $\psi_{\bm{k}n}(\bm{r})$ is the KS eigenfunction with wavevector $\bm{k}$ and band-index $n$ with the band energy $\varepsilon_{\bm{k}n}$ and $v^{\rm KS}_{\rm eff}(\bm{r})$ is the effective KS potential including the electron-nucleus, electron-electron (Hartree), nucleus-nucleus, and exchange-correlation potentials. 

In Eq.~(\ref{eq:KS}), $v_{U}(\bm{r})$ represents the DFT~+~$U$ potential, which is explicitly given by~\cite{Shick1999,Tran2008} 
\begin{align}
&v_{U}(\bm{r})
=\sum_{\nu}\sum_{mm'}\frac{\partial E_{U}}{\partial n_{mm'}^{\nu}}\frac{\delta n_{mm'}^{\nu}}{\delta \psi_{\bm{k}n}^{*}(\bm{r})}\\
&\quad\quad~=\sum_{\nu}\sum_{mm'}v_{mm'}^{\nu}\frac{\delta n_{mm'}^{\nu}}{\delta\psi_{\bm{k}n}^{*}(\bm{r})},\label{eq:vu}\\
&v_{mm'}^{\nu}=\delta_{mm'}\frac{U}{2}(1-n_{m}^{\nu})+(1-\delta_{mm'})(-Un_{mm'}^{\nu}),
\label{eq:vmm1}
\end{align}
where $\delta n_{mm'}^{\nu}/\delta \psi_{\bm{k}n}^{*}$ in Eq.~(\ref{eq:vu}) is the projector for $\psi_{\bm{k}n}$
and 1st (2nd) term in Eq.~(\ref{eq:vmm1}) is proportional to the diagonal (off-diagonal) density matrix $n_{mm}^{\nu}$~($n_{mm'}^{\nu}$). 

In self-consistent calculation, we use $\bm{k}$-mesh of 30$^2\times$21 corresponding to 1496 $\bm{k}$-points in the irreducible part of BZ for the space group $P4/nmm$ with the lattice constant $a$=$b$=$3.769{\rm \AA}$ and $c$=$5.521{\rm \AA}$, 
the internal coordinates $(\frac{x}{a},\frac{y}{a},\frac{z}{c})$=$(\frac{3}{4},\frac{1}{4},0)$ for Fe and $(\frac{1}{4},\frac{1}{4},z_{\rm Se})$ for Se with $z_{\rm Se}$=0.2688~\cite{McQueen2009}, and $R_{\rm MT}^{\rm Fe(Se)}=2.25~(2.14)$ bohr with the plane wave cutoff of $R_{\rm MT}K_{\rm max}=8$. 

In the calculation we restrict ourselves to zero expectation value of the magnetic moment. 
Although magnetic solutions have been also obtained for spin-polarized DFT and DFT~+~$U$ calculations~\cite{Subedi2008,Glasbrenner2015}, 
their magnetic moments are known to be overestimated. If one uses theories that incorporate strong correlation effects, such as DMFT, the discrepancy will be largely resolved and nonmagnetic solutions will be recovered.  
Therefore, the present nonmagnetic DFT~+~$U$ approach can be a good starting point for describing FeSe, which is nonmagnetic in experiment.

Here we note that qualitatively similar results have been obtained even if orbital-dependent interactions including Hund's coupling $J$ are used for $U$ (e.g., Fig.~4 in Ref.~\cite{Long2020}). Therefore, the main conclusion of the nematic state in the present study is expected to be unchanged even in the presence of Hund's coupling $J$.

In obtaining a symmetry broken nonmagnetic solution with WIEN2k, it is necessary to prepare an appropriate initial charge density obtained from the SCF converged result in the distorted lattice as $(x_{\rm Fe},y_{\rm Fe})\rightarrow (x_{\rm Fe}+\delta,y_{\rm Fe}+\delta)$ with $\delta=0.03$. In addition to this, the following points should be noted: select the space group $P1$ and directly input the coordinates of each atom in the unit cell in the structure file (case.struct), leaving the lattice constants and $R_{\rm MT}$ the same as in the original structure file. Set the local rotation matrix in the same file to the unit matrix, and set NUMBER OF SYMMETRY OPERATIONS to 1 to give the unit matrix. Skip the x symmetry and x group processes which are the default initial charge calculation.

\begin{figure*}[t]
\centering
\includegraphics[width=17.5cm]{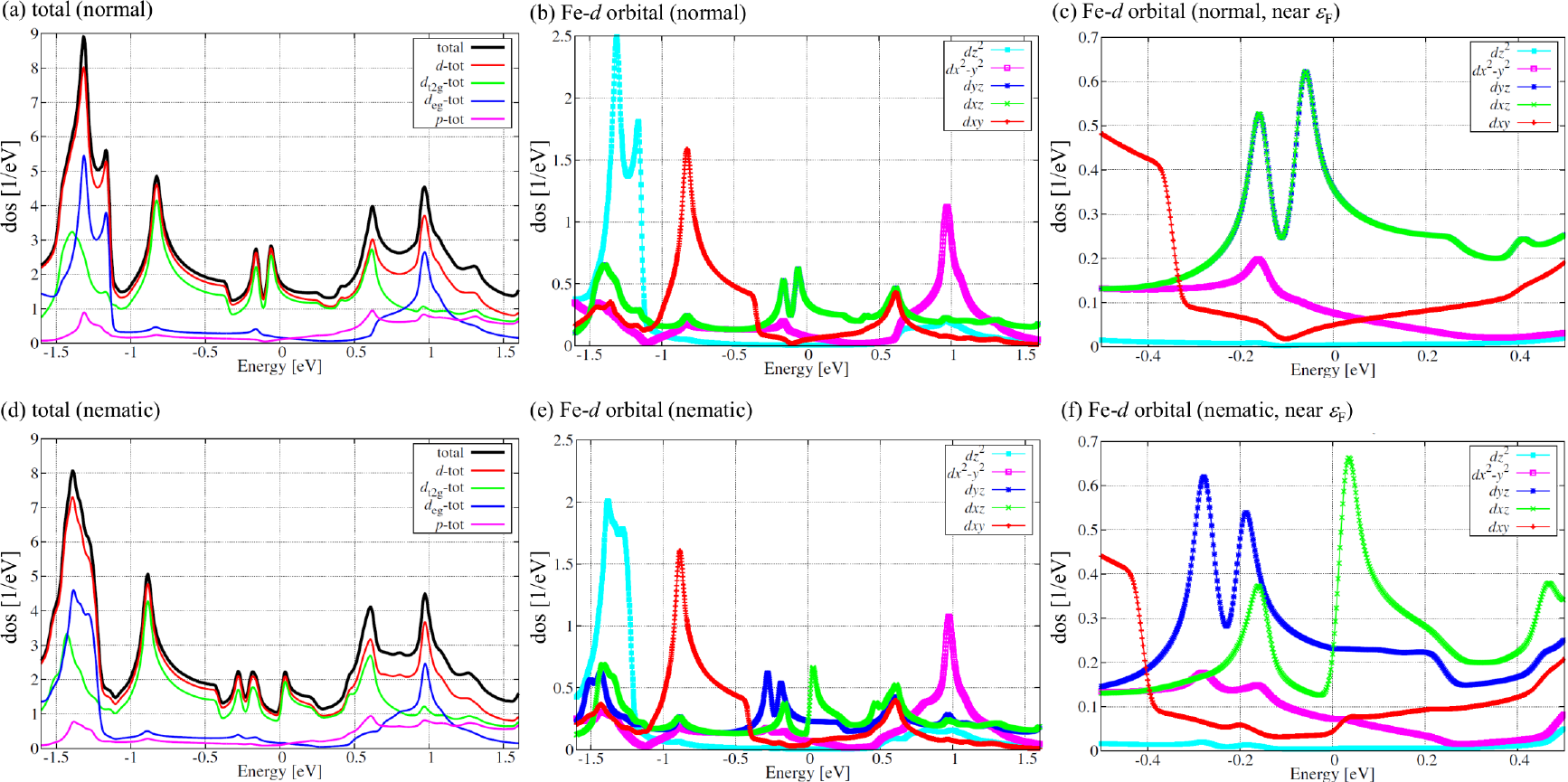}
\caption{(Color online)
DOS for (a)$-$(c) normal state ($U=3.3$~eV) and (d)$-$(f) nematic state ($U=3.4$~eV) 
with 
(a) and (d) total, Fe-$d$ total, $dt_{2g}$-total, $de_g$-total and $p$-total DOS per atom and 
(b) and (e) $d$-orbital components with color plot of 
$d_{xy}$~(red), 
$d_{xz}$~(green), 
$d_{yz}$~(blue), 
$d_{x^2-y^2}$~(pink), 
$d_{z^2}$~(light blue). 
The panels of (c),(f) are the same plot of (b),(e) with low energy range $-0.6\sim 0.6$~eV. 
}\label{Fig-s1}
\end{figure*}

\section{Density-of-states of normal and nematic states}
In order to investigate the electronic states in correspondence with the band structure in Fig.~4 in the main text, we show the density of states (DOS) in the normal ($U=3.3$~eV) and nematic ($U=3.4$~eV) states in Fig.~\ref{Fig-s1}. 
Figures~\ref{Fig-s1}(a) and S1(d) demonstrate that the electronic states around $\varepsilon_{\rm F}$(=0 eV) originate from the Fe-$d$ orbitals in both the normal and nematic states, which is basically consistent with the results of FeSe DFT calculations~\cite{Subedi2008,Lohani2015}. 
In addition, the $d_{z^2}$ and $d_{x^2-y^2}$ orbitals have large peaks at $-1.3$~eV and 1~eV, respectively, indicating that the low-energy states are dominated by the $d_{xy}$, $d_{xz}$, and $d_{yz}$ orbitals, as shown in Figs.~\ref{Fig-s1}(b) and \ref{Fig-s1}(d). 
A comparison of Figs.~\ref{Fig-s1}(c) and \ref{Fig-s1} (f) shows that the $d_{xz}$ and $d_{yz}$ orbitals, which are degenerate in the normal state, show a remarkable splitting in the nematic state. 
The spectral shape of the DOS of the $d_{xz}$ orbital shifting to high-energy region as compared with the normal state suggests that the shift is not rigid-band-like. Furthermore, DOS of the $d_{xy}$ and $d_{x^2-y^2}$ orbitals changes with the normal-nematic transition as a consequence of hybridization with the $d_{xz}$/$d_{yz}$ orbitals as described in the main text.


\end{document}